\documentclass[letter]{aa} 

\usepackage{hyperref}
\usepackage{graphicx}
\usepackage{txfonts}
\begin{document}


   \title{Early emission lines in SN 2024ggi revealed by high-resolution spectroscopy}



   \author{Thallis Pessi
          \inst{1}
          \and
          Régis Cartier
          \inst{1}
          \and
          Emilio Hueichapan
          \inst{1}
          \and
          Danielle de Brito Silva
          \inst{1,2}
          \and
          Jose L. Prieto
          \inst{1,3}
          \and
          Ricardo R. Muñoz
          \inst{4}
          \and
          Gustavo E. Medina
          \inst{5}
          \and
          Paula Diaz
          \inst{4}
          \and 
          Ting S. Li
          \inst{5}
          }

    \institute{Instituto de Estudios Astrof\'isicos, Facultad de Ingenier\'ia y Ciencias, Universidad Diego Portales, Av. Ej\'ercito Libertador 441, Santiago, Chile\\ 
              \email{thallis.pessi@mail.udp.cl}
         \and 
         Millenium Nucleus ERIS
         \and
            Millennium Institute of Astrophysics MAS, Nuncio Monse\~nor Sotero Sanz 100, Off. 104, Providencia, Santiago, Chile
        \and
        Departamento de Astronomía, Universidad de Chile, Camino El Observatorio 1515, Las Condes, Santiago, Chile
        \and
        Department of Astronomy and Astrophysics, University of Toronto, 50 St. George Street, Toronto, ON M5S 3H4, Canada
        }

   \date{}

 
  \abstract
   {We present an analysis of very early high-resolution spectroscopic observations of the  {Type II} supernova (SN) 2024ggi, a nearby SN that occurred in the galaxy NGC 3621 at a distance of $7.24$~Mpc ($z \approx 0.002435$). These observations represent the earliest high-resolution {spectra} of a {Type II SN} ever made.}
   {We analyzed the very early-phase spectroscopic evolution of SN 2024ggi obtained in a short interval at $20.6$ and $27.8$~h after its discovery, or $26.6$ and $33.8$~h after the SN first light. Observations were obtained with the high-resolution spectrograph MIKE ($R \approx 22 \ 600 - 28 \ 000 $) at the $6.5$~m \textit{Magellan Clay} Telescope, located at the Las Campanas Observatory, on the night of April 12, 2024 UT.}
   {{The emission lines were identified and studied in detail during the first hours of SN 2024ggi.}
   We analyzed the evolution of ions of {\ion{H}{I}, \ion{He}{I}, \ion{He}{II}, \ion{N}{III}, \ion{C}{III}, \ion{Si}{IV}, \ion{N}{IV,} and \ion{C}{IV} detected across the spectra. We modeled these features with multiple Gaussian and Lorentzian profiles, and estimated their velocities and full widths at half maximum (FWHMs).}}
   { {The spectra show asymmetric emission lines of \ion{H}{I}, \ion{He}{II}, \ion{C}{IV}, and \ion{N}{IV} that can be described by narrow  Gaussian cores  (FWHM $\leq 200$\,km\,s$^{-1}$) with broader Lorentzian wings, and symmetric narrow emission lines of \ion{He}{I}, \ion{N}{III}, and \ion{C}{III}. 
   The emission lines of \ion{He}{I} are detected only in the first spectrum, indicating the rapid ionization of \ion{He}{I} to \ion{He}{II}.
   The narrow components of the emission lines show a systematic blueshift relative to their zero-velocity position, with an increase of $\sim18$\,km\,s$^{-1}$ in the average velocity between the two epochs.  
   The broad Lorentzian components show a blueshift in velocity relative to the narrow components, and a significant increase in the average velocity of $\sim103$\,km\,s$^{-1}$. 
   Such a rapid evolution and significant ionization changes in a short period of time were never observed before, and are probably a consequence of the radiative acceleration generated in the SN explosion.}}
   {}


   \maketitle
%

\section{Introduction} \label{sec:intro}

\begin{figure*}[t!]
\centerline{\includegraphics[scale=0.65]{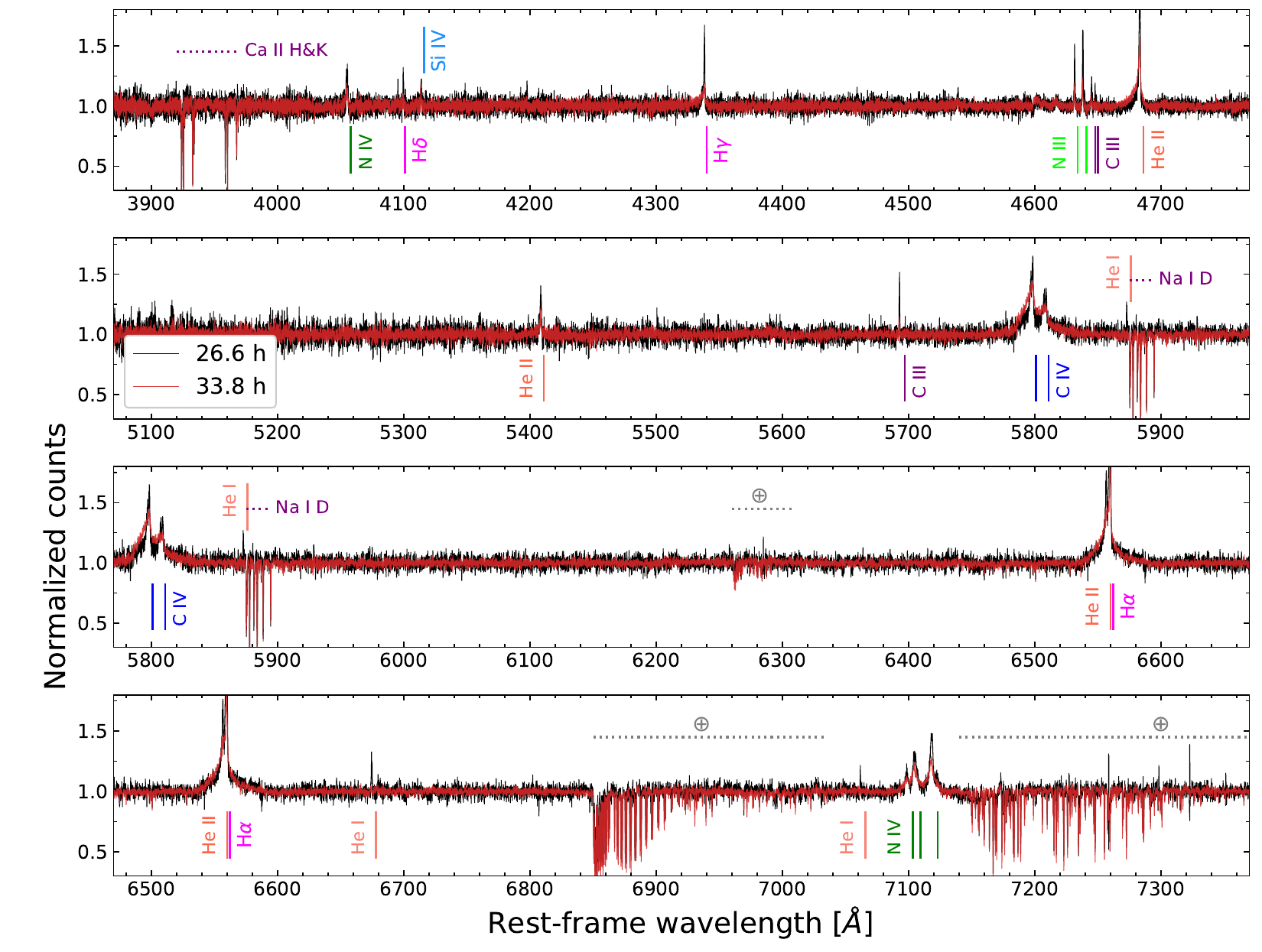}}
\caption{Complete normalized MIKE/Magellan Clay spectra of SN 2024ggi. The black and red spectra were obtained, respectively, at $26.6$~h and $33.8$~h after the SN's first light. Vertical lines mark the main emission features in the spectra, horizontal dotted gray lines mark the telluric absorption features, and horizontal dotted purple lines mark the \ion{Na}{I} D and \ion{Ca}{II} H\&K absorption lines. The spectra show strong emission lines of {\ion{H}{I}}, \ion{He}{I}, \ion{He}{II}, \ion{N}{III}, \ion{N}{IV}, \ion{C}{III}, and \ion{C}{IV}. The top panel shows the spectra obtained by the blue arm of the spectrograph, and the three other panels show the spectra obtained along the red arm. \label{fig:red_arm}} 
\end{figure*}

The discovery of nearby and young {Type II} supernovae (SNe) provides a singular opportunity
to study these events in detail. The close proximity allows for very detailed observations to be obtained, from the early explosion to the late nebular phases. 
Observations of early {Type II SN} phases allow for a characterization of the initial explosion and the last moments in the life of their massive progenitor stars. A characteristic feature of these early moments is the appearance of narrow emission lines (usually called flash-ionization lines) generated by the photoionization of dense circumstellar material (CSM) close to the progenitor star \citep[e.g.,][]{2014Natur.509..471G,2016ApJ...818....3K,2017NatPh..13..510Y,2017A&A...605A..83D,2021ApJ...912...46B,2023ApJ...952..119B}. Analyses of these early features can constrain the chemical composition, temperature, and density of the CSM, as well as properties of the progenitor star \citep[e.g.,][]{2014A&A...572L..11G, 2020MNRAS.496.1325B, 2022ApJ...926...20T, 2022ApJ...924...15J, 2024arXiv240302382J}.  

The recent discovery of the nearby SN 2023ixf \citep[on May 19, 2023;][]{2023TNSTR1158....1I} in the galaxy M101, at a distance of $6.85$~Mpc \citep{2022ApJ...934L...7R}, was followed by extensive analyses of its evolution \citep[e.g.,][]{2023ApJ...953L..16H, 2023ApJ...954L..12T, 2023ApJ...955L...8H, 2023PASJ...75L..27Y}. The classification spectrum of SN 2023ixf showed features of H, He, C, and N \citep{2023TNSCR1164....1P}. Further analyses of the early spectral evolution revealed highly ionized species of \ion{He}{I}, \ion{He}{II}, \ion{C}{IV}, \ion{N}{III}, \ion{N}{IV}, and \ion{N}{V}, and helped to place constraints {on} the progenitor and CSM properties \citep[e.g.,][]{2023ApJ...954L..42J, 2023ApJ...956...46S, 2023ApJ...956L...5B}.

SN 2024ggi was discovered extremely early by the Asteroid Terrestrial-impact Last Alert System (ATLAS) survey \citep{2018PASP..130f4505T} at 2024-04-11.14 UT (MJD = 60411.14), at an apparent magnitude of $o \approx 18.92$~mag \citep{2024TNSTR1020....1T, 2024TNSAN.100....1S, 2024arXiv240609270C}, and was soon reported by other dedicated surveys \citep{2024TNSAN.101....1K, 2024TNSAN.102....1C, 2024TNSAN.108....1K}. Early spectroscopic classification confirmed it as a Type II SN with {emission lines with narrow cores and broad and asymmetric wings} \citep{2024TNSAN.103....1H, 2024TNSAN.104....1Z}, and {the potential progenitor star was identified as a red supergiant of $\sim13$~M$_\odot$ in pre-explosion images} \citep[e.g.,][]{2024TNSAN.105....1Y, 2024TNSAN.107....1P}.
Recent analyses of the early evolution of SN 2024ggi were presented by \citet{2024arXiv240419006J}, \citet{2024arXiv240518490S}, \citet{2024arXiv240609270C}, and \citet{2024arXiv240607806Z}.
The SN is located at RA = 11:18:22.087, Dec = $-32$:50:15.27 (coordinates taken from the last object coordinate update of the Transient Name Server\footnote{\url{https://www.wis-tns.org/object/2024ggi}}, made on April 18, 2024), in one of the southeast spiral arms of the galaxy NGC 3621, at a projected distance of $3.87$~kpc from the galaxy center (see Fig. \ref{fig:ngc3621}). 
The Cepheid distance to NGC 3621 is {$7.24\pm0.20 $~Mpc} \citep{2006ApJS..165..108S}, as reported in the NASA/IPAC Extragalactic Database\footnote{\url{https://ned.ipac.caltech.edu/byname?objname=NGC+3621}}.

Here, we present an analysis of the very early spectroscopic evolution of SN 2024ggi, obtained at $20.6$ and $27.8$~h after its discovery ($26.6$ and $33.8$~h after the SN's first light), with the high-resolution \textit{Magellan Inamori Kyocera} Echelle (MIKE) spectrograph at the \textit{Magellan Clay} Telescope \citep{2003SPIE.4841.1694B}. This Letter is organized as follows: In Sect. \ref{sec:obs} we describe our observations and data reduction. In Sect. \ref{sec:res} we report constraints on the early-phase spectroscopic features of SN 2024ggi and a comparison with other early observations of core-collapse SNe. Finally, in Sect. \ref{sec:conc} we present a summary and our conclusions.

\section{Observations} \label{sec:obs}

Observations of SN 2024ggi were conducted with the double echelle MIKE instrument at the $6.5$~m \textit{Magellan Clay }Telescope, located at the Las Campanas Observatory. The spectra were obtained on the night of April 12, 2024, at 00:22:01 UT (MJD = 60412.015) and 07:04:24 UT (MJD = 60412.294), respectively $20.6$~h and $27.8$~h after its discovery. 
Observations of the two spectra were taken with a $1.0\arcsec \times 5.0 \arcsec $ slit, a $2 \times 2$ binning mode, and at an airmass of $1.15$ and $1.81$, respectively.
The MIKE spectra have wavelength resolutions of $R \approx 22 \ 600$ for the red side and $R \approx 28 \ 000$ for the blue side (measured at the calibration lamp spectra).
The spectral regions considered in this analysis are from $3570 \ \AA$ to $4780 \ \AA$ in the blue arm and from $4900 \ \AA$ to $7400 \ \AA$ in the red arm. The spectral regions below $3570 \ \AA$ and between $4780 \ \AA$ and $4900 \ \AA$ have a low signal-to-noise ratio, and no clear spectral features are detected. The spectral regions above $7400 \ \AA$, in the red arm, are dominated by telluric lines.

The data were reduced using the standard  \texttt{Carnegie Python Mike pipeline} \citep{kelson2000evolution,kelson2003optimal}. The pipeline automatically performs wavelength calibration and flat-fielding, and provides as a result a spectrum divided into different orders. With the aim of creating the complete 1D spectrum with all the orders combined, we carefully inspected the wavelength calibration of the output spectra, checking the consistency between different orders in their overlapping regions and using the position of narrow lines as a sanity check. The 1D spectra (in units of counts) were carefully normalized using a low-order polynomial to fit the continuum, with particular attention paid to the normalization at the edge of each order. In the fitting process we avoided regions that included narrow or broad spectral features. Once normalized, we joined the different orders, computing the mean of the normalized flux weighted by its uncertainty in the overlapping regions. After the red and blue arms were normalized, the barycentric correction was performed. This corrected for the Doppler shift introduced by the rotation of the Earth around the Sun, thus placing the spectra in the heliocentric system of reference. 

We used the publicly available early ATLAS photometry to estimate the time of first light of SN 2024ggi, and report the methodology in Appendix \ref{app:time}. We find that the time of the first light was at $t_{0} = 60410.89 \pm 0.14$ days, or $6$~h before the time of discovery, and adopt this value throughout this work. 
From the early ATLAS light curve, we estimate that the SN had an apparent magnitude of $o \approx 14.4$~{mag} at the time of the first observation, and of $o \approx 13.7$~{mag} during the second observation.

\begin{figure*}[h!]
\centering{
\includegraphics[scale=0.32]{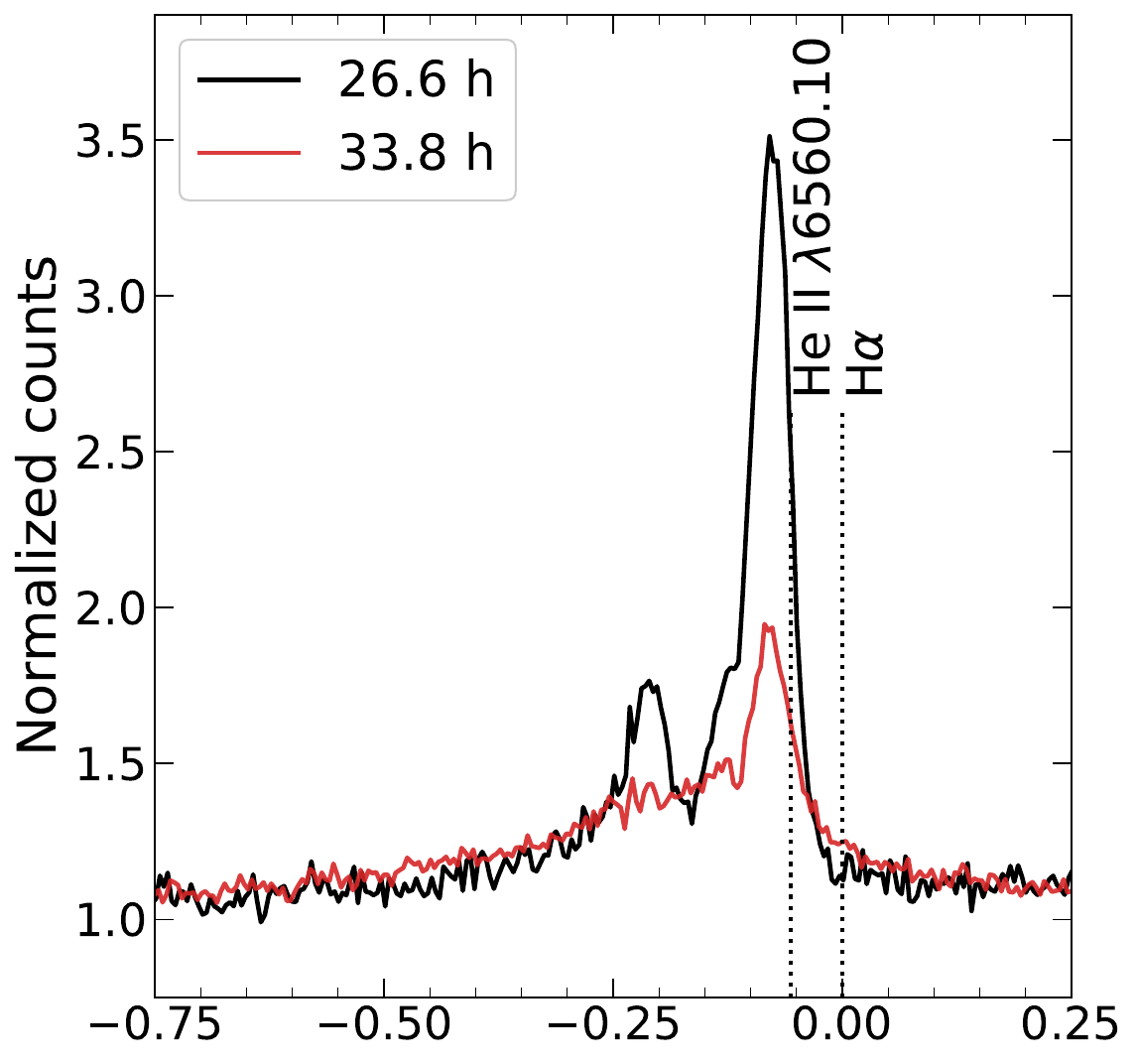}
\includegraphics[scale=0.32]{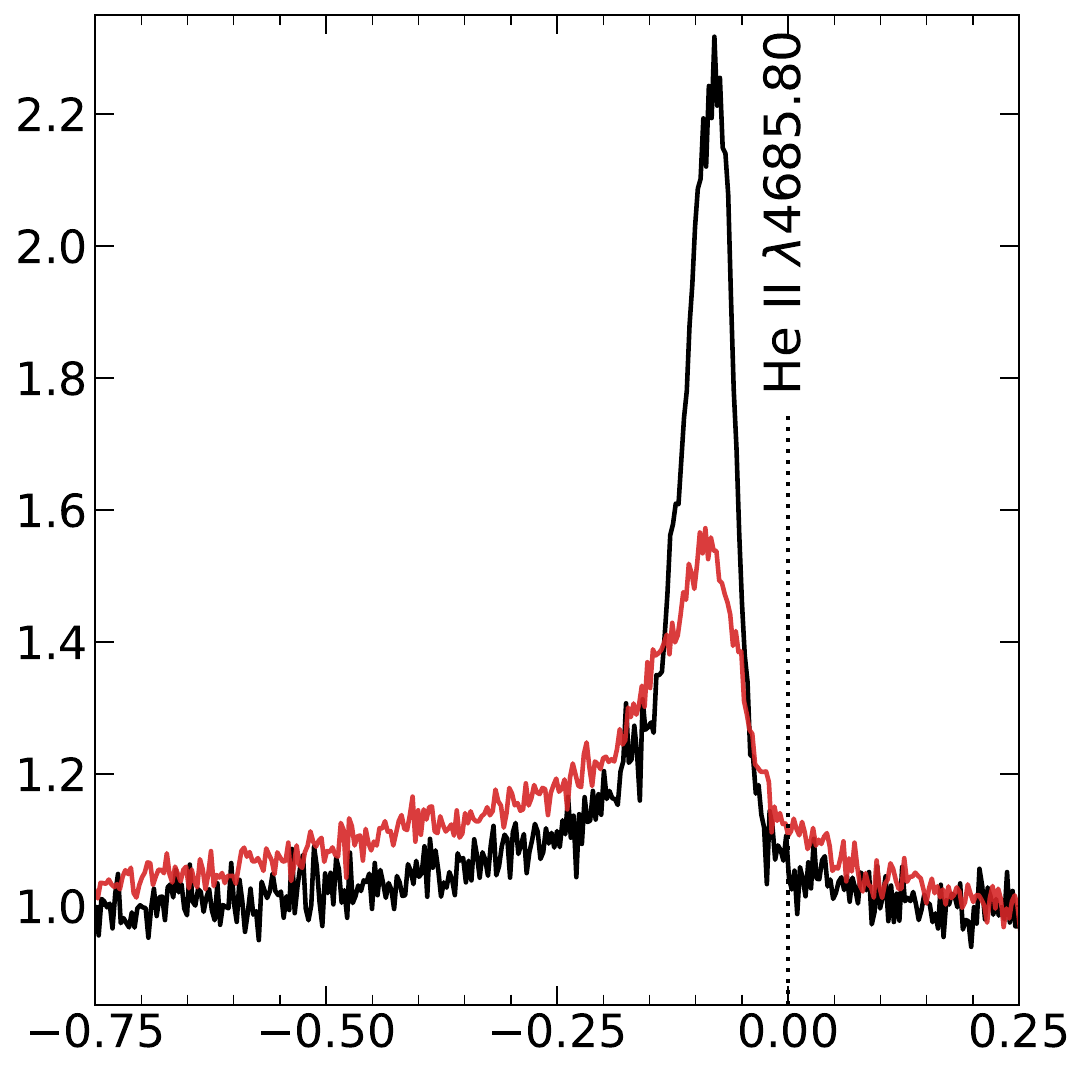}
\includegraphics[scale=0.32]{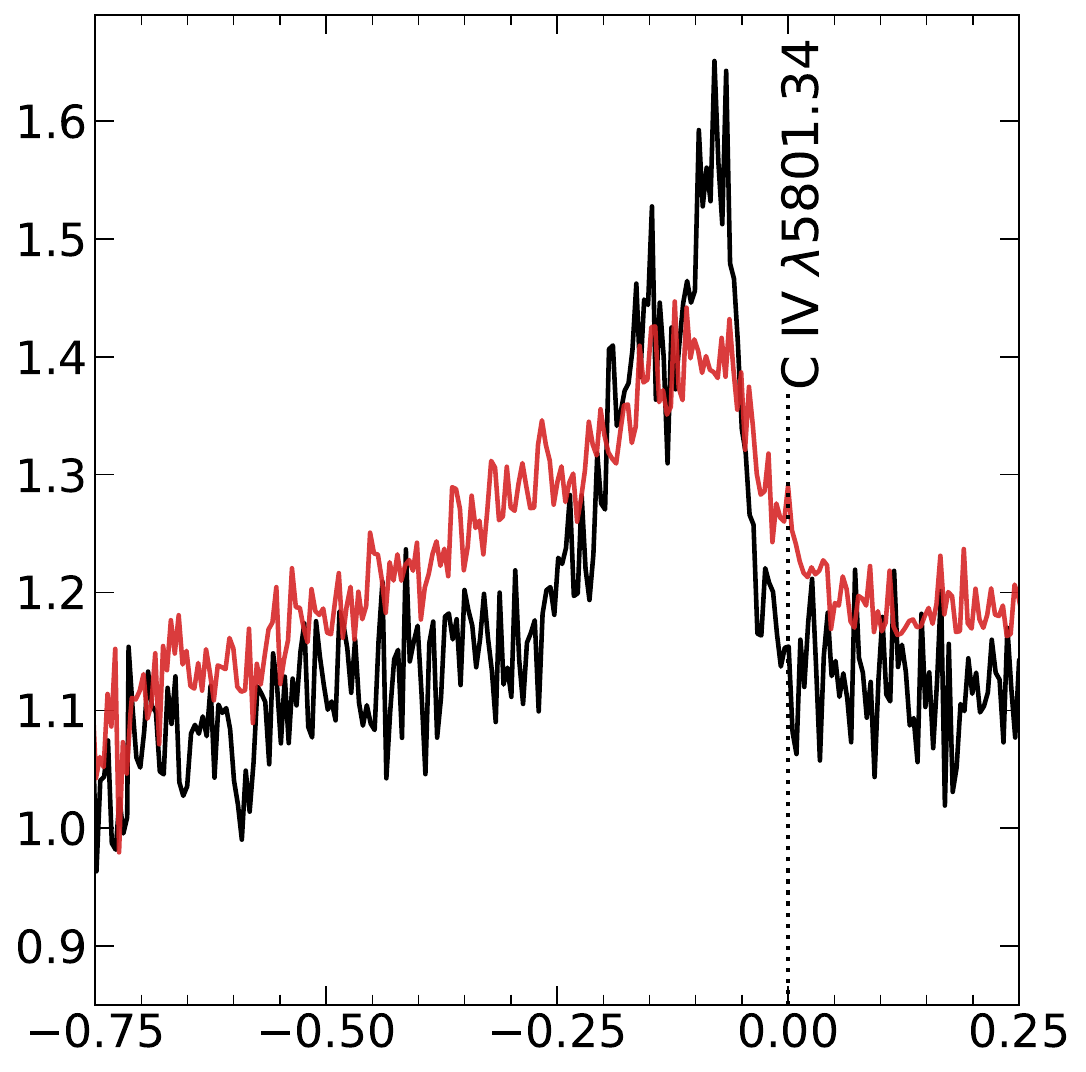}
\includegraphics[scale=0.32]{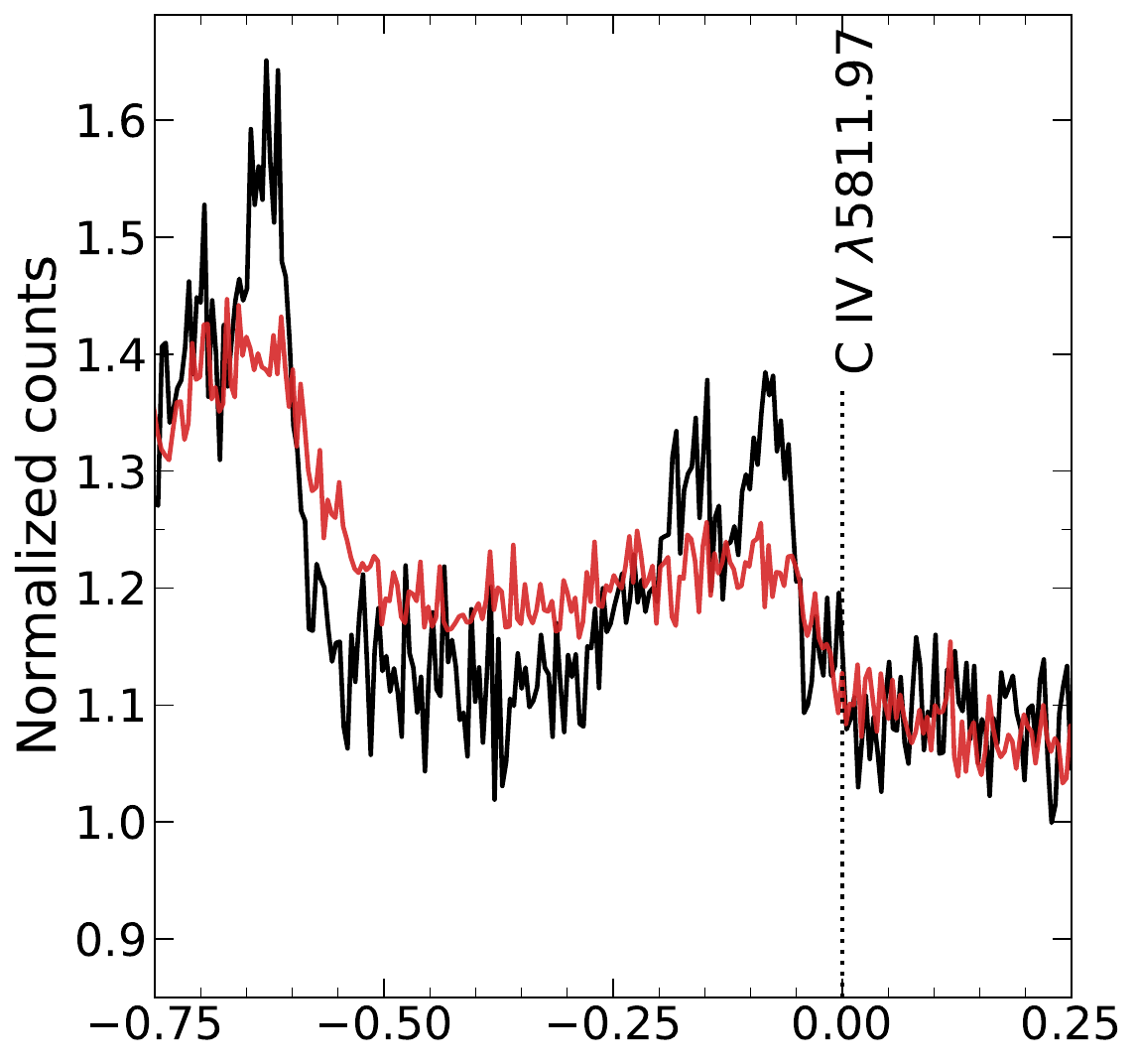}
\includegraphics[scale=0.32]{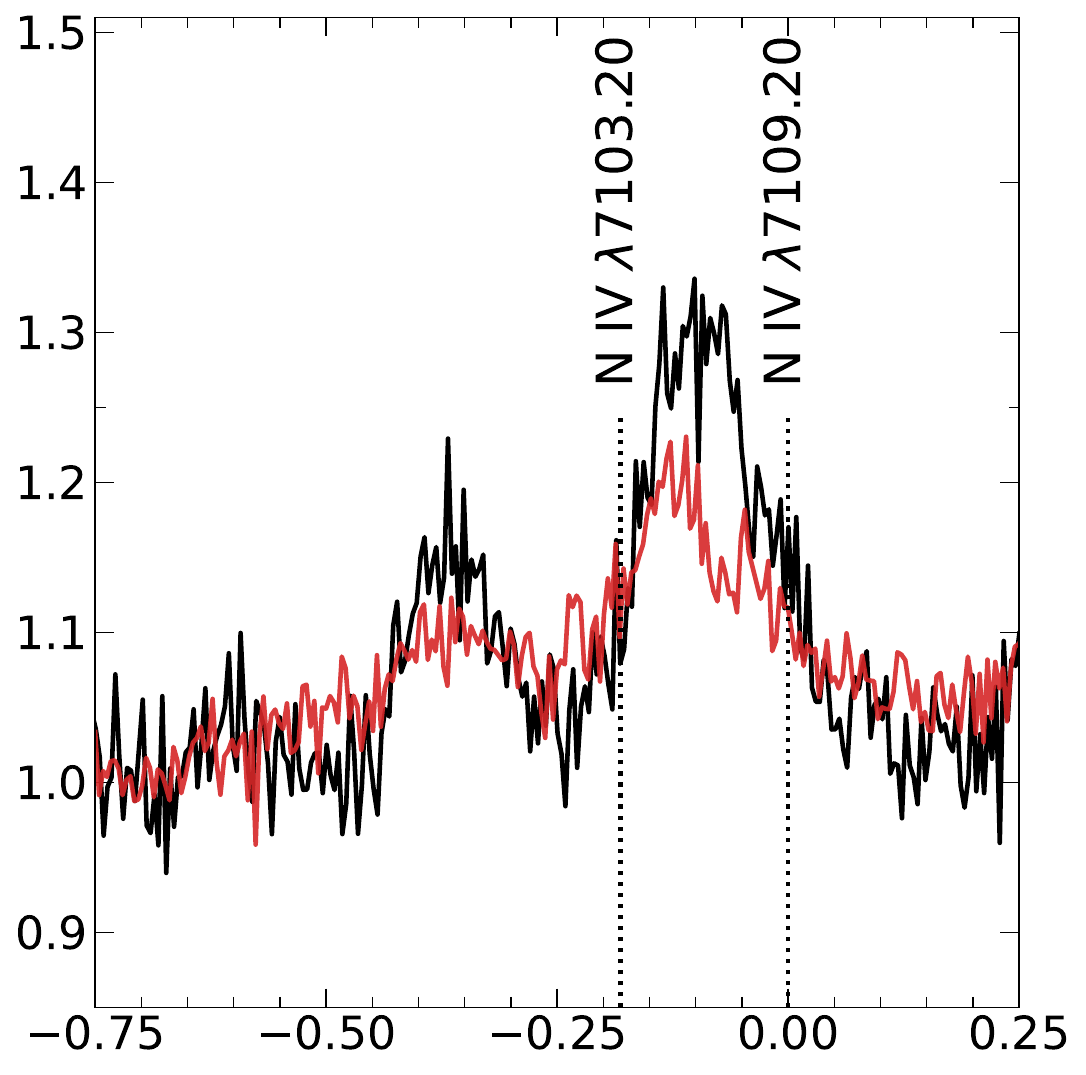}
\includegraphics[scale=0.32]{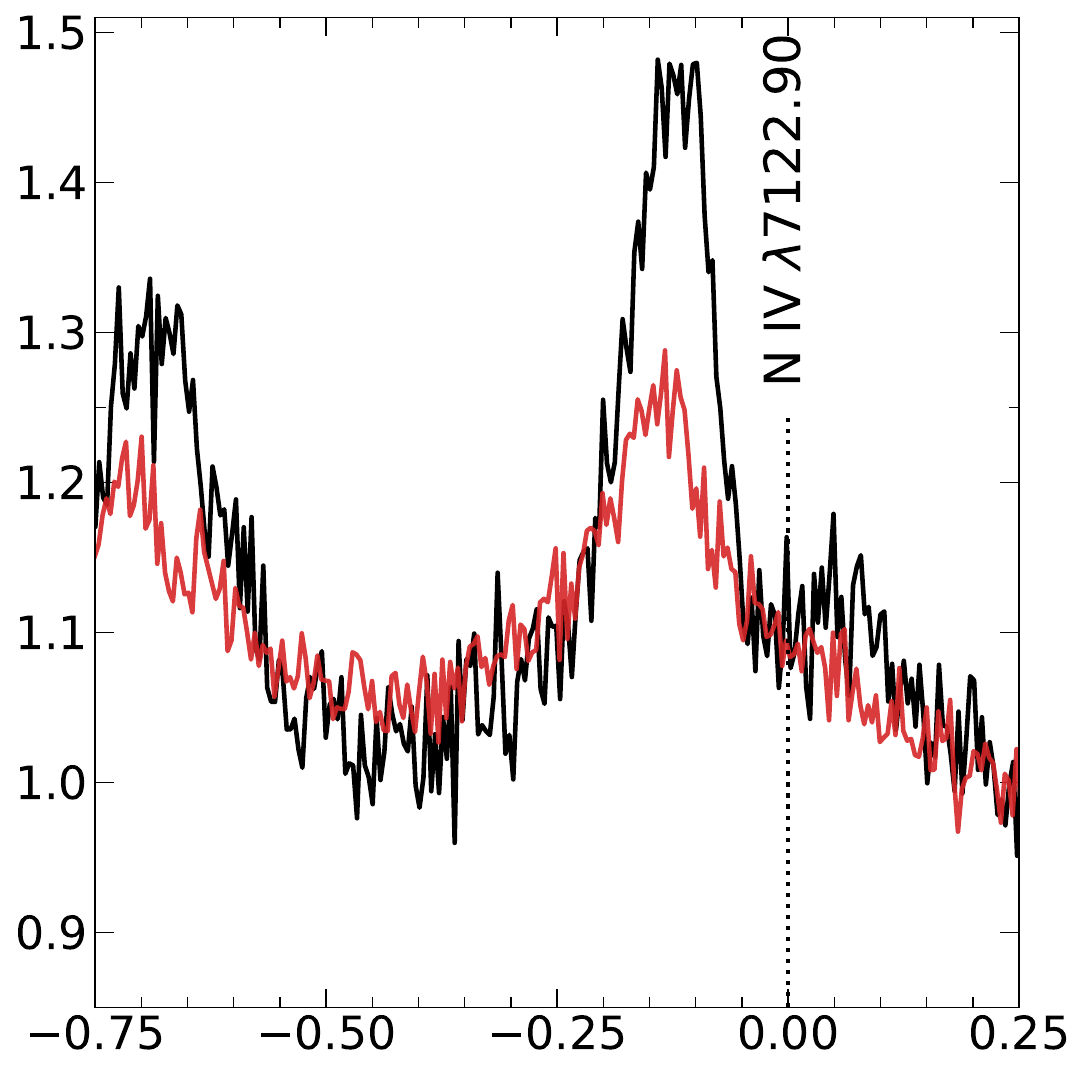}
\includegraphics[scale=0.32]{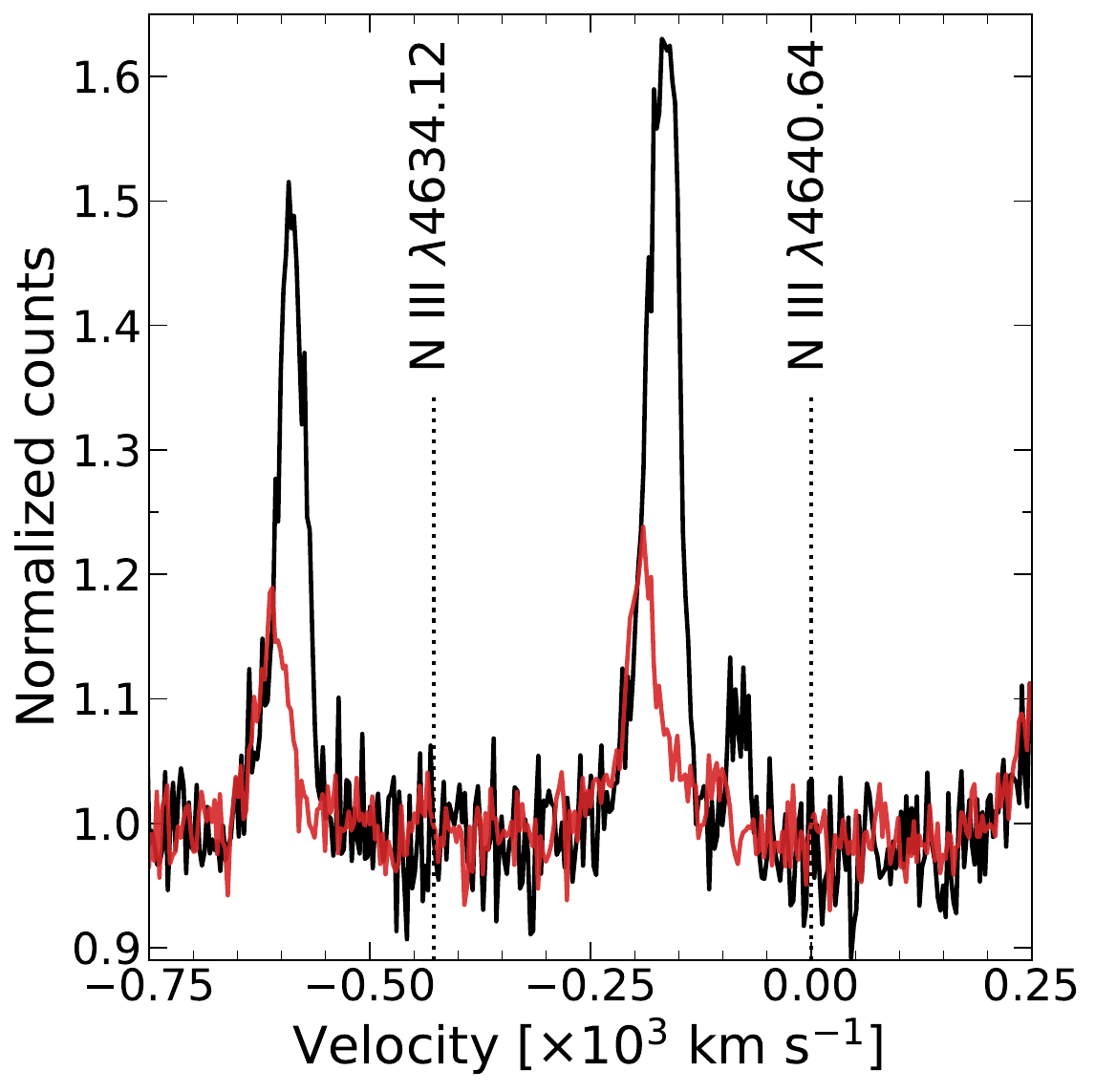}
\includegraphics[scale=0.32]{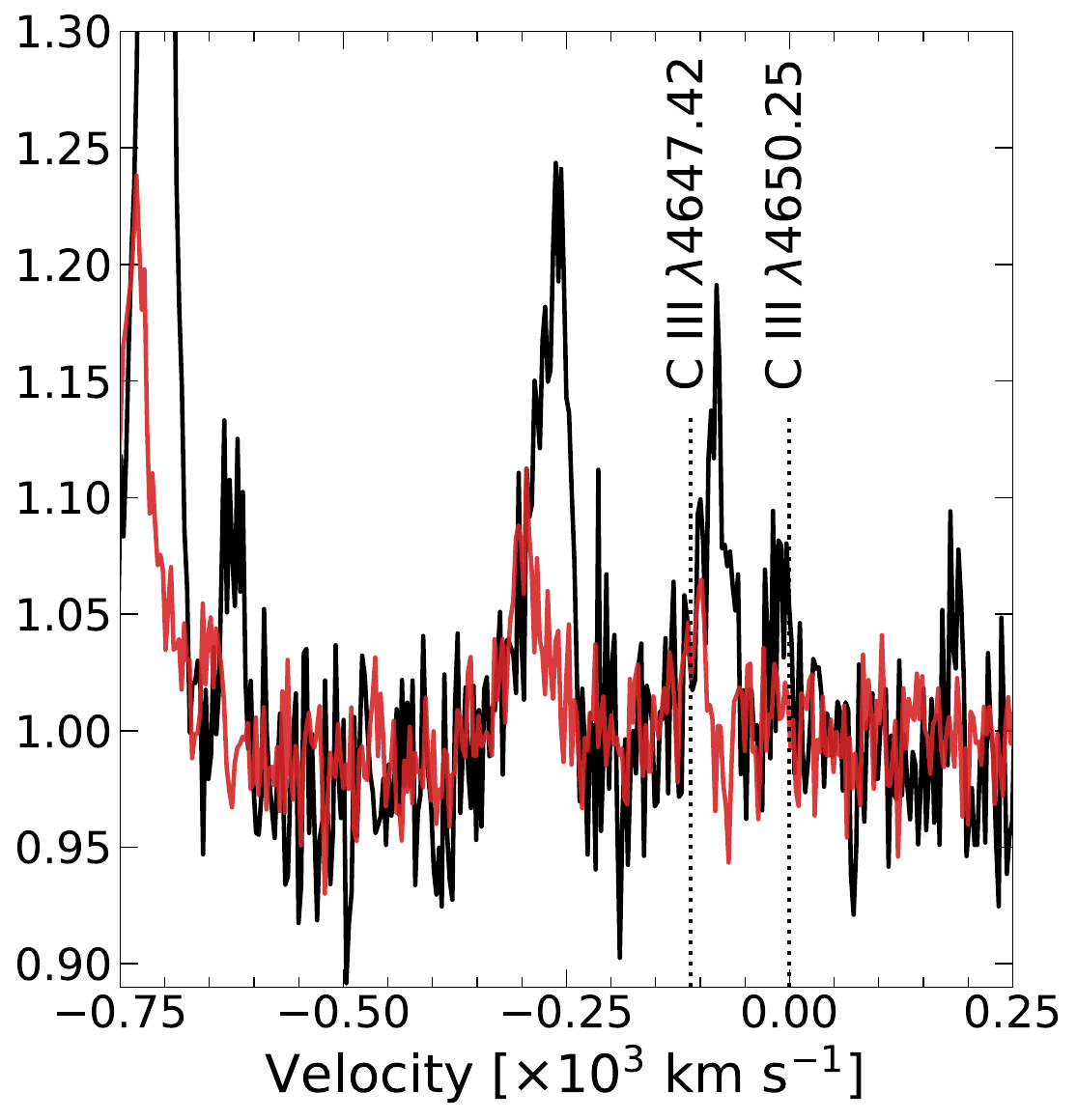}
\includegraphics[scale=0.32]{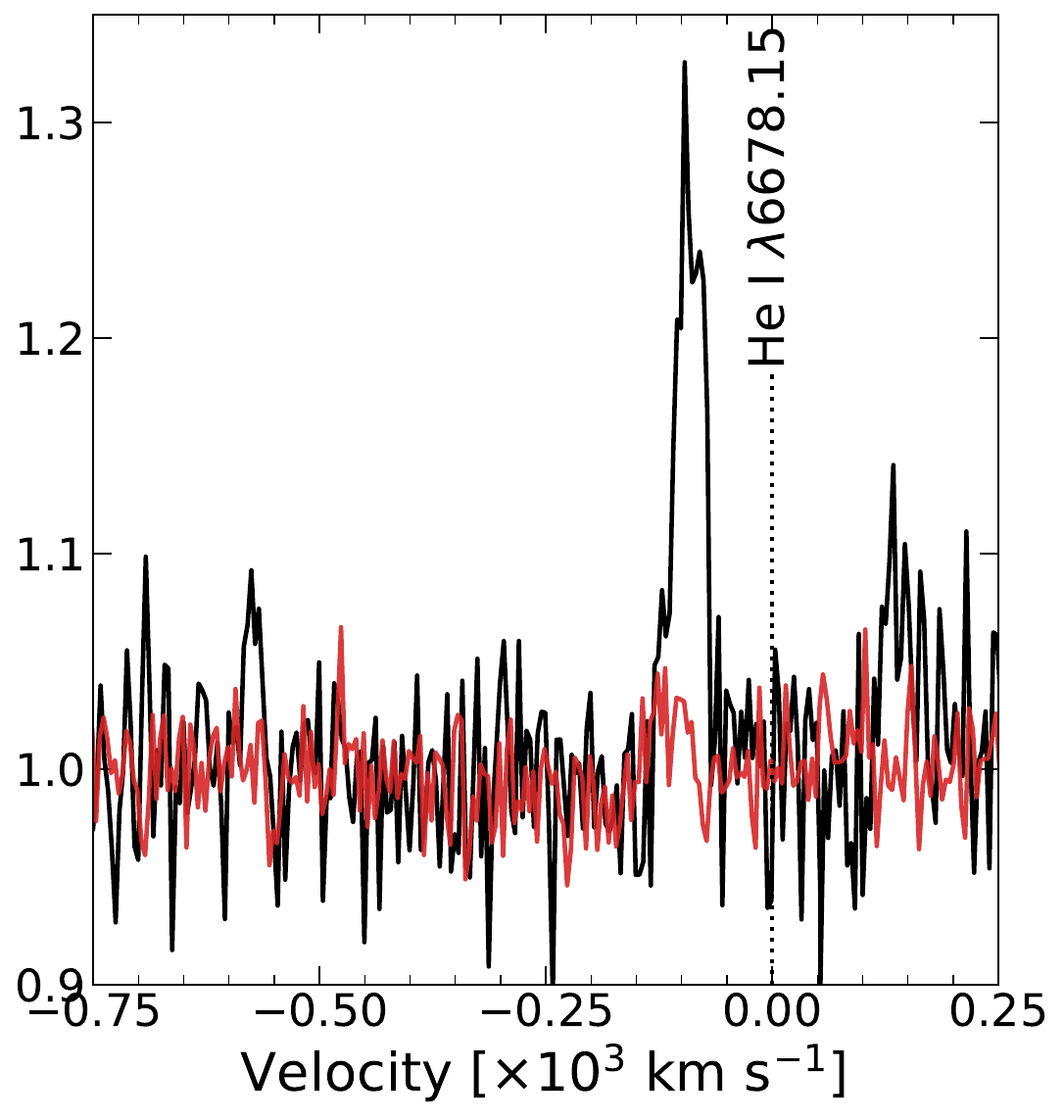}
}
\caption{{Evolution of the  H$\alpha$, \ion{He}{II} $\lambda6560.10$, \ion{He}{II} $\lambda4685.80$, \ion{C}{IV} $\lambda\lambda5801.34, 5811.97$, \ion{N}{IV} $\lambda\lambda\lambda7103.20, 7109.20, 7122.90$, \ion{N}{III} $\lambda\lambda4634.12, 4640.64$, \ion{C}{III} $\lambda\lambda4647.42, 4650.25$, and \ion{He}{I} $\lambda6678.15$ lines, at $26.6$~h (in black) and $33.8$~h (in red) after the first light. 
Spectra are shifted to the zero-velocity positions of H$\alpha$, \ion{He}{II} $\lambda4685.80$, \ion{C}{IV} $\lambda5801.34$, \ion{C}{IV} $\lambda5811.97$, \ion{N}{IV} $\lambda7109.20$, \ion{N}{IV} $\lambda7122.90$, \ion{N}{III} $\lambda4640.64$, \ion{C}{III} $\lambda4650.25$, and \ion{He}{I} $\lambda6678.15$. }
\label{fig:CIV}} 
\end{figure*}

The spectra presented here were corrected using the heliocentric recession velocity of NGC 3621 of $730 \ \textrm{km} \ \textrm{s}^{-1}$ \citep[$z \approx 0.002435$][]{2004AJ....128...16K}. Figure \ref{fig:NaI} shows the detail of the \ion{Na}{I} D and \ion{Ca}{II} H\&K absorption lines detected in the spectra. {Using these features, we measured a velocity blueshift of $-71.7 \pm 0.4$\,km\,s$^{-1}$ of the center position of these absorption features relative to the heliocentric velocity of NGC 3621, which we adopted as the projected rotation velocity of the host at the SN location, corresponding to an effective redshift correction of $z=0.002196\pm0.000001$. We applied this correction to the spectra shown in Fig. \ref{fig:CIV} and used it throughout our analysis.
Additionally, we measure a total Galactic extinction in the line of sight of SN 2024ggi of $E(B-V)_{\textrm{Gal}} = 0.12 \pm 0.02$~mag and a host galaxy extinction of $E(B-V)_{\textrm{host}} = 0.036 \pm 0.007$~mag, resulting in a total extinction of $E(B-V)_{\textrm{total}} = 0.16 \pm 0.02$ mag (see our methodology in Appendix \ref{app:na}).
}

\section{Results and discussion} \label{sec:res}

\subsection{Properties of the emission features}

Figure \ref{fig:red_arm} shows the complete high-resolution MIKE spectra of SN 2024ggi, at $26.6$~h (in black) and $33.8$~h (in red) after the time of first light, normalized to the continuum counts. Vertical lines mark the positions of the main emission features identified in the spectra. 
Telluric absorption features, \ion{Ca}{II} H\&K, and the \ion{Na}{I} D absorption lines are also highlighted. 
Both spectra show emission lines of {\ion{H}{I}}, \ion{He}{I}, \ion{He}{II}, \ion{N}{III}, \ion{C}{III}, {\ion{Si}{IV}}, \ion{N}{IV}, \ion{and C}{IV}\footnote{These emission features are also observed in low-resolution spectra obtained at similar phases \citep[e.g.,][]{2024arXiv240419006J, 2024arXiv240607806Z}.}. 
We identify \ion{He}{I} $\lambda5875.62$, $\lambda6678.15$, and $\lambda7065.18$, \ion{He}{II} $\lambda4685.80$, $\lambda5411.53$, and $\lambda6560.10$, the \ion{N}{III} $\lambda\lambda4634.12, 4640.64$ and \ion{C}{III} $\lambda\lambda4647.42, 4650.25$ doublets in the blue part of the spectra, and an emission line of \ion{C}{III} $\lambda5695.92$.
The spectra also show strong \ion{C}{IV} $\lambda\lambda5801.34, 5811.97$ doublet emission features, emission from the \ion{N}{IV} $\lambda\lambda\lambda7103.20, 7109.20, 7122.90$ triplet, and \ion{N}{IV} $\lambda4057.75$ and {\ion{Si}{IV} $\lambda4116.10$} emission lines.

{We fit the main emission lines in the spectra with multiple Gaussian and Lorentzian profiles, and report their central position relative to the zero velocity and full width at half maximum (FWHM) in Tables \ref{tab:gaussian_lines} and \ref{tab:lor_lines}.}
{Figure \ref{fig:CIV} shows the region around the emission lines of  H$\alpha$, \ion{He}{I} $\lambda6678.15$, \ion{He}{II} $\lambda4685.80$, \ion{C}{IV} $\lambda\lambda5801.34, 5811.97$, \ion{N}{IV} $\lambda\lambda\lambda7103.20, 7109.20, 7122.90$, \ion{N}{III} $\lambda\lambda4634.12, 4640.64$, \ion{C}{III} $\lambda\lambda4647.42, 4650.25$, and \ion{He}{II} $\lambda6560.10$, where we see their evolution between $26.6$~h and $33.8$~h after the first light. }

{The H$\alpha$, \ion{He}{II}, \ion{C}{IV}, and \ion{N}{IV} emission lines are asymmetric and can be described by narrow Gaussian cores (FWHM $\leq 200$\,km\,s$^{-1}$) with broader Lorentzian wings.
These line profiles are generated by the narrow recombination lines being diffused by electron-scattering, creating broader wings in the inner and denser regions of the CSM. Because the ionizing radiation is higher in the inner regions, the broader Lorentzian features are only observed for the higher ionization lines in the spectra \citep[these features are commonly observed in Type IIn SNe; see, e.g.,][]{2009MNRAS.394...21D, 2013MNRAS.431.2599M, 2015ApJ...806..213S}.
H$\alpha$ also shows a weak P Cygni absorption in the second epoch, at $\sim -114.7$\,km\,s$^{-1}$.
Figure \ref{fig:CIV_gaussians} shows the different components used to model the \ion{C}{IV} feature at the two epochs. 
Although H$\alpha$, \ion{He}{II}, and \ion{N}{IV} are fit with single Gaussians and Lorentzians, we fit the \ion{C}{IV} lines in the first epoch with a combination of two Gaussians and one Lorentzian profile. This reproduces the observed double peaks in the \ion{C}{IV} emission lines well. 
It is possible that the double-peak profile is due to an asymmetric density distribution of the CSM producing shells with similar levels of ionization but different levels of radiative acceleration. An alternative explanation is the light travel effect between the near and rear sides of a heterogeneous CSM, also indicative of some degree of inhomogeneity in the CSM distribution.}

{The narrow components of most of the detected emission lines are characterized by a systematic blueshift. On average, at the first epoch the lines are blueshifted by $-93.3$\,km\,s$^{-1}$, with a dispersion of $15.02$\,km\,s$^{-1}$. At the second epoch, we measure an average blueshift velocity of $-111.6$\,km\,s$^{-1}$, with a similar dispersion. H$\alpha$ and \ion{C}{IV} are exceptions, not showing any significant blueshift of their narrow component between the two epochs (although, in the first epoch, \ion{C}{IV} shows a peculiar double-peaked narrow component, which, fit with a single Gaussian profile, yields a more blueshifted velocity). The velocities of the narrow Gaussian components are reported in Table \ref{tab:gaussian_lines}.}
{Assuming that the CSM is optically thick, we used the velocity of the narrow component of H$\alpha$ to estimate the velocity of the wind, and obtain $v_{wind} \approx 77$\,km\,s$^{-1}$.}

{The broad Lorentzian components of \ion{H}{I}, \ion{He}{II}, \ion{C}{IV}, and \ion{N}{IV} show a blueshift not only in relation to their zero-velocity position, but also in relation to their narrow components \citep[as observed in, e.g., SN 1998S;][]{2015ApJ...806..213S}, which generates the observed asymmetries in the wings (see Table \ref{tab:lor_lines} and Fig. \ref{fig:CIV_gaussians}).
We measure the average velocity of the Lorentzian profiles to be $-154.3$\,km\,s$^{-1}$ in the first spectrum and $-257.6$\,km\,s$^{-1}$ in the second.
This increase in velocity could potentially be explained by the radiative acceleration of the unshocked CSM due to the large radiative flux diffusing through the optically thick inner CSM \citep[see, e.g.,][]{2002MNRAS.330..473C, 2023ApJ...952..115T}.}

{The emission lines of \ion{N}{III} $\lambda\lambda4634.12, 4640.64$ and \ion{C}{III} $\lambda\lambda4647.42, 4650.25$ are relatively symmetric, showing only subtle deviations from Gaussian profiles in the $26.6$~h after the first light.
The \ion{C}{III} lines show a stronger peak in the red part of the emission profile, reminiscent of the double-peaked \ion{C}{IV} line at the same epoch, while the \ion{N}{III} lines may have a weak tail extending into the blue. 
These potential asymmetries of \ion{C}{III} and \ion{N}{III} are very subtle, and the best fits are obtained using a single Gaussian rather than a more complex combination of profiles.}
{Similar to the narrow components of the other emission lines, \ion{N}{III} and \ion{C}{III} show an increase in velocity between the two epochs. The blueshift of the peak of these lines can be seen clearly in Fig. \ref{fig:CIV}. }{We measure the center of the Gaussian profile fit to the \ion{N}{III} $\lambda4640.64$ line in the first spectrum at $-96.4\pm0.5$\,km\,s$^{-1}$ and in the second spectrum at $-120.3\pm0.9$\,km\,s$^{-1}$.} 
{The \ion{N}{III} and \ion{C}{III} are probably being generated in the outer parts of the CSM, as they are described only by narrow components and have a lower ionization than their \ion{N}{IV} and \ion{C}{IV} counterparts. Thus, this increase in velocity might have the same origin as the narrow components of the other emission lines.}

{Finally, we detect a narrow and symmetric emission line of \ion{He}{I} $\lambda6678.15$ in the first spectrum. This line becomes fainter in the second epoch, detected only with a low signal-to-noise ratio (S/N $\sim 1.6$).}
{The other observed emission lines of \ion{He}{I} also seem to disappear in the interval of time between the two spectra, indicating that a large fraction of the \ion{He}{I} in the CSM is rapidly ionized to \ion{He}{II}. 
This is likely also a consequence of radiative acceleration suffered by the ions and electrons in the CSM due to the increase in the radiation field that also completely ionizes the helium in the CSM, suppressing the \ion{He}{I} lines.
The \ion{He}{I} lines detected with a good significance ($\sigma\textgreater3$) are well modeled by a symmetric single Gaussian profile, with blueshifted velocities in the range $75-92$\,km\,s$^{-1}$ and FWHMs of $23-36$\,km\,s$^{-1}$.

\begin{figure}[t!]
\centerline{\includegraphics[scale=0.45]{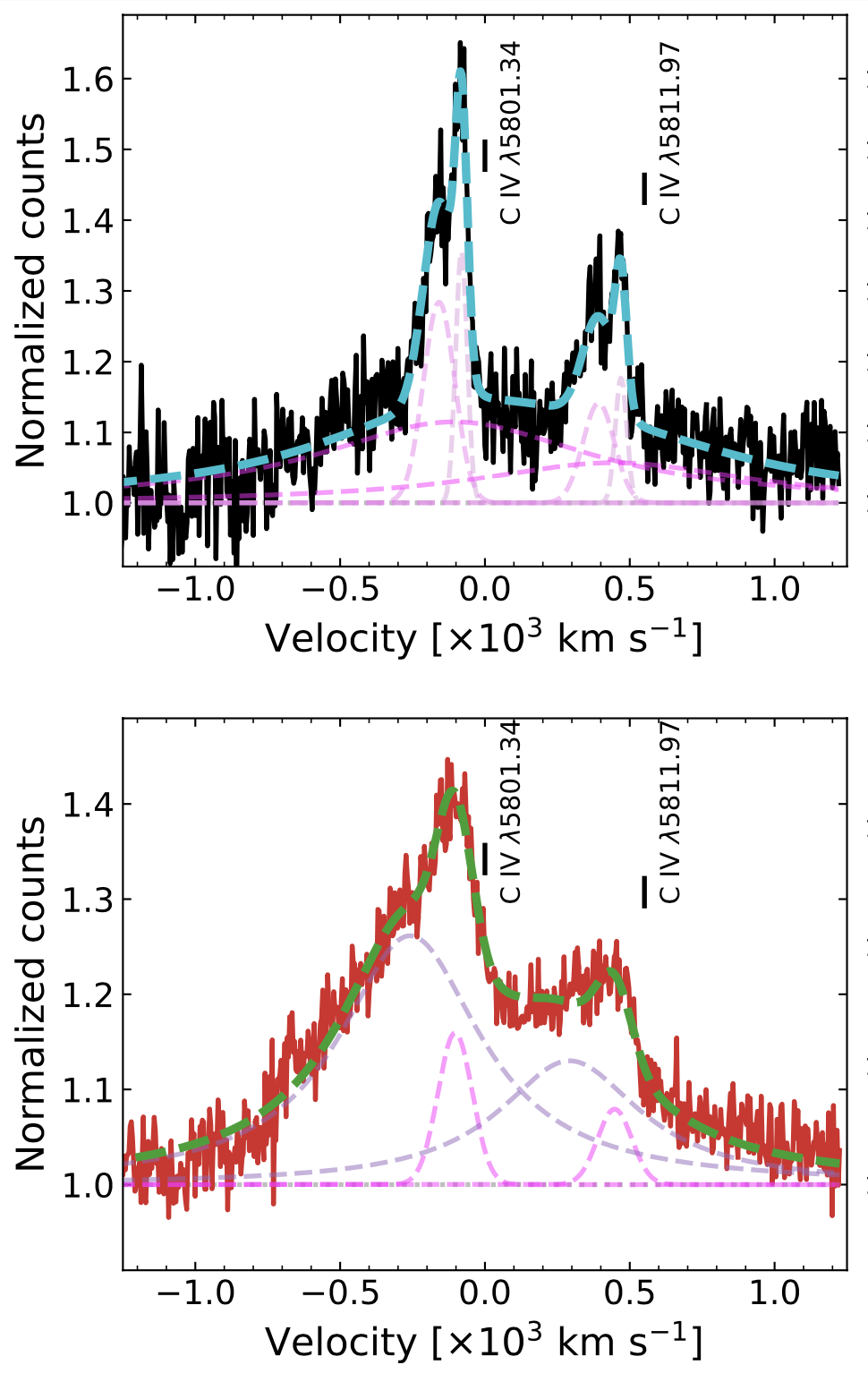}}
\caption{{Narrow Gaussian and broad Lorentzian profiles used to model the emission lines of \ion{C}{IV} $\lambda\lambda5801.34, 5811.97$. The top and bottom panel show, respectively, the lines at $26.6$~h and $33.8$~h after the first light. The narrow components of the \ion{C}{IV} lines can be modeled by a combination of two Gaussian profiles, which describes the observed double peaks well.} \label{fig:CIV_gaussians}} 
\end{figure}

\subsection{Comparison with the literature}

The early spectra of SN 2024ggi presented here show similar emission lines as the early high-resolution spectra of SN 2023ixf taken at $2.6 - 7.6$~d after the explosion and analyzed in \citet{2023ApJ...956...46S}. 
We identify {emission} lines of {\ion{H}{I}, \ion{He}{I}}, \ion{He}{II}, \ion{N}{III}, \ion{C}{III}, \ion{N}{IV}, and \ion{C}{IV} present in both SNe, with the {\ion{H}{I}, \ion{He}{II}, \ion{N}{IV}, and \ion{C}{IV}} having narrow Gaussian cores with broader Lorentzian components in both cases. 
The narrow components have similar FWHMs of $<200  \ \textrm{km} \ \textrm{s}^{-1}$, with the broad wings of \ion{He}{II} $\lambda4685.80$ for SN 2023ixf also extending up from $\sim -500 \ \textrm{km} \ \textrm{s}^{-1}$ to $\sim -1000 \ \textrm{km} \ \textrm{s}^{-1}$ during its earliest epochs. However, the \ion{C}{IV} lines in SN 2023ixf seem to have broader features than SN 2024ggi, with the broad wings extending up to $\sim -2000 \ \textrm{km} \ \textrm{s}^{-1}$.
{Unlike SN 2024ggi, SN 2023ixf does not show absorption P Cygni features from the unshocked CSM in the early spectra, indicating that there is no emission toward the line of sight. \citet{2023ApJ...956...46S} interpreted this as being generated by an asymmetric CSM around the SN.
The narrow emission line components of SN 2023ixf also show a systematic blueshift of $\sim -115 \ \textrm{km} \ \textrm{s}^{-1}$, which is slightly higher than the velocity we measure for SN 2024ggi.}
Finally, similar to \citet{2023ApJ...956...46S}, we do not detect the broad emission feature near $4600 \ \AA$ observed in other Type II SNe at early times \citep[e.g.,][]{2007ApJ...666.1093Q, 2018MNRAS.476.1497B, 2018ApJ...861...63H, 2023ApJ...945..107P}.

{
Emission lines described by a combination of narrow cores with broader Lorentzian wings, generated by the electron-scattering of diffused recombination lines, are commonly observed for Type IIn SNe \citep[e.g.,][]{2009MNRAS.394...21D, 2013MNRAS.431.2599M, 2014ApJ...797..118F, 2015ApJ...806..213S}. \citet{2015ApJ...806..213S} describe similar emission line shapes observed in the early high-resolution spectra of SN 1998S. The broad Lorentzian features in SN 1998S also show a systematic blueshift relative to their narrow components, which were interpreted by \citet{2015ApJ...806..213S} as being a consequence of radiative acceleration \citep[also see][]{2002MNRAS.330..473C}. We find a similar effect for SN 2024ggi (with the added advantage of having two spectra taken in a short interval of time), whose broad emission line components experience a significant change in velocity.}

\begin{table*}
\small
\caption{Central velocities, FWHMs, and line positions from Gaussian profiles fitted to the emission lines in the spectra. \label{tab:gaussian_lines}}
\renewcommand{\arraystretch}{1.4}
\centering
\begin{tabular}{lcccccc}
\hline
Ion & Velocity$_{a}$ & Velocity$_{b}$ & FWHM$_{a}$ & FWHM$_{b}$ & Wavelength$_{a}$ & Wavelength$_{b}$ \\
& [km s$^{-1}$] & [km s$^{-1}$] & [km s$^{-1}$] & [km s$^{-1}$] & [$\AA$] & [$\AA$] \\
\hline
\ion{Si}{IV} $\lambda 4116.1$ & $-99.3\pm2.4$ & $-122.7\pm2.3$ & $55.9\pm6.4$ & $48.6\pm6.9$ & $4114.73 \pm 0.03$ & $4114.42 \pm 0.03$ \\
\ion{N}{III} $\lambda 4634.1$ & $-97.7\pm0.5$ & $-119.1\pm0.9$ & $36.9\pm1.5$ & $39.9\pm2.0$ & $4632.63 \pm 0.01$ & $4632.29 \pm 0.01$ \\
\ion{N}{III} $\lambda 4640.6$ & $-96.4\pm0.5$ & $-120.3\pm0.9$ & $41.9\pm1.2$ & $44.9\pm3.8$ & $4639.15 \pm 0.01$ & $4638.78 \pm 0.01$ \\
\ion{C}{III} $\lambda 4647.4$ & $-91.1\pm1.8$ & $-120.5\pm2.0$ & $43.1\pm4.1$ & $42.6\pm5.9$ & $4646.01 \pm 0.03$ & $4645.55 \pm 0.03$ \\
\ion{C}{III} $\lambda 4650.2$ & $-90.4\pm2.1$ & $-112.1\pm3.9^{*}$ & $35.6\pm6.2$ & $22.4\pm7.7^{*}$ & $4648.85 \pm 0.03$ & $4648.51 \pm 0.06^{*}$ \\
\ion{C}{III} $\lambda 5995.9$ & $-97.3\pm0.6$ & $-109.8\pm2.1$ & $27.3\pm1.6$ & $29.4\pm6.4$ & $5694.07\pm 0.01$ & $5693.84\pm 0.04$ \\
\ion{He}{II} $\lambda 4685.7$  & $-89.3\pm0.6$ & $-101.9\pm0.9$ & $55.9\pm1.5$ & $112.3\pm3.7$ & $4684.30\pm 0.01$ & $4684.11 \pm 0.02$ \\
\ion{He}{II} $\lambda 5411.5$ & $-93.6\pm2.3$ & $-120.5\pm4.5$ & $56.0\pm5.4$ & $84.6\pm34.6$ & $5409.83\pm 0.04$ & $5409.35 \pm 0.08$ \\
\ion{He}{II} $\lambda 6560.1$ & $-90.2\pm1.5$ & ...            & $32.8\pm2.7$ & ...           & $6558.13\pm 0.03$ & \\ 
H$\gamma$                    & $-85.4\pm1.0 $ & $-99.8\pm3.9$ & $48.9\pm3.5$ & $71.2\pm13.7$ & $4339.23\pm 0.01$ & $4339.02\pm 0.06$ \\
H$\alpha$ & $-77.9\pm0.3$ & $-77.8\pm0.8$ & $45.9\pm0.8$ & $47.6\pm2.7$ & $6561.09\pm0.01$ & $6561.09\pm0.01$ \\
\ion{He}{I} $\lambda 3888.6$ & $-62.0\pm5.7^{*}$ & ... & $60.2\pm10.9^{*}$ & ... & $3887.79\pm 0.07^{*}$  & ... \\
\ion{He}{I} $\lambda 4471.5$ & $-62.4\pm9.1^{*}$ & ... & $27.3\pm17.5^{*}$ & ... & $4469.91\pm 0.06^{*}$  & ... \\
\ion{He}{I} $\lambda 5875.9$ & $-85.9\pm1.8$ & ... & $23.7\pm3.2$  & ... & $5874.21\pm 0.04$  & ... \\
\ion{He}{I} $\lambda 6678.2$ & $-91.9\pm1.2$ & $-116.0\pm7.0^{*}$ & $36.4\pm2.6$  & $38.8\pm10.6^{*}$ & $6676.10\pm 0.03 $  & $6675.56\pm 0.16^{*}$ \\
\ion{He}{I} $\lambda 7065.2$ & $-74.6\pm1.8$ & ... & $23.3\pm3.9$  & ... & $7063.41\pm 0.04$  & ... \\
\ion{C}{IV}  $\lambda 5801.3$ & $-81.1\pm3.1$ & $-102.9\pm3.2$ & $47.1\pm7.9$ & $139.8\pm7.9$ & $5799.74\pm1.06$ & $5799.31\pm0.06$ \\
$^{\dagger}$\ion{C}{IV} $\lambda 5801.3$ & $-158.2\pm9.3$ & ... & $128.3\pm29.2$ & ... & $5798.25\pm0.18$ & ... \\
\ion{N}{IV} $\lambda 7123.0$ & $-127.6\pm10.9$ & $-143.9\pm83.3$ & $98.0\pm23.2$ & $166.8\pm89.1$ & $7119.95\pm 0.26$  & $7119.56\pm1.9$ \\
\ion{N}{IV} $\lambda 4057.8$ & $-134.9\pm1.9$ & $-157.3\pm3.8$ & $102.1\pm3.8$ & $ 134.1\pm7.3$ & $44055.93\pm 0.03$  & $4055.63\pm 0.05$ \\
\hline
\end{tabular}
\tablefoot{Labels \textit{a} and \textit{b} correspond to the spectra at the first and second epochs, respectively. $^{*}$Weak emission line, detected to less than $3 \sigma$.$^{\dagger}$Values for the high-velocity Gaussian fit of the \ion{C}{IV} profile for the first spectrum.}
\end{table*}

\begin{table*}
\small
\caption{Central velocities, FWHMs, and line positions from Lorentzian profiles fit to the emission lines in the spectra. \label{tab:lor_lines}}
\renewcommand{\arraystretch}{1.4}
\centering
\begin{tabular}{lcccccc}
\hline
Ion & Velocity$_{a}$ & Velocity$_{b}$ & FWHM$_{a}$ & FWHM$_{b}$ & Wavelength$_{a}$ & Wavelength$_{b}$\\
& [km s$^{-1}$] & [km s$^{-1}$] & [km s$^{-1}$] & [km s$^{-1}$] & [$\AA$] & [$\AA$] \\
\hline
H$\gamma$                     & $-164.9\pm14.6$ & $-218.6\pm12.9$ & $146.7\pm11.8$ & $330.8\pm23.2$ & $4338.08\pm 0.21$ & $4337.30\pm 0.19$ \\
H$\alpha$                     & $-165.5\pm4.3$  & $-177.3\pm4.5$  & $319.9\pm15.6$ & $452.7\pm9.0$  & $6559.18\pm0.09$ & $6558.92\pm0.06$ \\
\ion{He}{II} $\lambda 4685.7$ & $-151.9\pm6.1$  & $-272.6\pm8.8$   & $136.4\pm10.0$ & $373.5\pm10.6$  & $4683.33\pm0.09$ & $4681.44\pm 0.14$ \\
\ion{He}{II} $\lambda 5411.5$ & $-159.4\pm13.9$ & $-265.0\pm118.3$ & $45.6\pm17.1$  & $340.5\pm132.9$ & $5408.64\pm0.25$ & $5406.74\pm 2.13$ \\
\ion{C}{IV}  $\lambda 5801.3$ & $-107.2\pm55.3$ & $-258.6\pm5.6$ & $1160.6\pm202.4$ & $595.5\pm13.8$ & $5799.24\pm1.07$ & $5796.31\pm0.12$ \\
\ion{N}{IV}  $\lambda 7123.0$ & $-167.4\pm20.0$ & $-338.4\pm87.9$ & $228.2\pm70.9$ & $347.2\pm57.9$ & $7119.01\pm0.48$ & $7114.94\pm2.09$ \\
\hline
\end{tabular}
\tablefoot{Labels \textit{a} and \textit{b} correspond to the spectra at the first and second epochs, respectively.}
\end{table*}

\section{Summary and conclusions} \label{sec:conc}

In this Letter {we have presented high-resolution spectroscopy of the nearby Type II SN 2024ggi, with two epochs obtained with the MIKE spectrograph at the \textit{Magellan Clay} Telescope, at $26.6$ and $33.8$~h after the SN first light.}
We have shown that the spectra are marked by asymmetric emission lines of \ion{H}{I}, \ion{He}{II}, \ion{C}{IV}, and \ion{N}{IV} that can be described by narrow Gaussian cores  (FWHM $\leq 200$\,km\,s$^{-1}$) with broader Lorentzian wings, {with the H$\alpha$ showing a weak P Cygni absorption in the second epoch}, and symmetric narrow emission lines of \ion{He}{I}, \ion{N}{III}, and \ion{C}{III}.

{We have shown that the narrow emission features are marked by a systematic blueshift offset from their rest wavelength, with an increase of $18.3$\,km\,s$^{-1}$ in the average blueshift velocity, between the first and second epochs. 
We have also observed a blueshift of the broad Lorentzian components, with a significant increase in the average velocity between the two epochs of $103.3$\,km\,s$^{-1}$. 
A significant increase in velocity in such a short period of time (we believe) has never before been observed.
This acceleration is probably generated by the radiation pressure of the SN emission interacting with the slower inner wind, and could explain the mismatch often found between the velocities observed in the narrow lines of young Type II SNe (a few hundred\,km\,s$^{-1}$) and the expected velocities of red supergiant winds (a few tens of \,km\,s$^{-1}$)}. 

Narrow emission lines of \ion{He}{I}  are detected in the first spectrum but disappear in the later epoch, indicating that a large fraction (or all) of \ion{He}{I} was rapidly ionized to \ion{He}{II}.
This is also a consequence of radiative acceleration that increases the radiation field and completely ionizes the helium material in the CSM.\


The spectra presented here represent the earliest high-resolution spectroscopic observations of a {Type II SN} ever made.
The short period between the two observations, of around $8$~h, allowed us to detect the rapid evolution and rapid ionization changes in the CSM located close to {Type II SNe} at early times.} 
Fast triggering and a higher cadence of early observations for future events are therefore extremely important.
Further observations of SN 2024ggi are strongly encouraged, as well as modeling of the CSM and the SN progenitor star using the data presented here.

\begin{acknowledgements}

T.P. acknowledges the support by ANID through the Beca Doctorado Nacional 202221222222.
J.L.P. acknowledges support from ANID, Millennium Science Initiative, AIM23-0001. D.D.B.S. acknowledges the support by Millennium Nucleus ERIS NCN2021\_017. R.R.M. gratefully acknowledges support by the ANID BASAL project FB210003. G.E.M. acknowledges support from the University of Toronto Arts \& Science Postdoctoral Fellowship program.


\end{acknowledgements}

\bibliography{ref}
\bibliographystyle{aasjournal}

\begin{appendix} 

\section{SN 2024ggi at NGC 3621}\label{app:ngc3621}

\begin{figure}[t!]
\centerline{\includegraphics[width=0.55\textwidth]{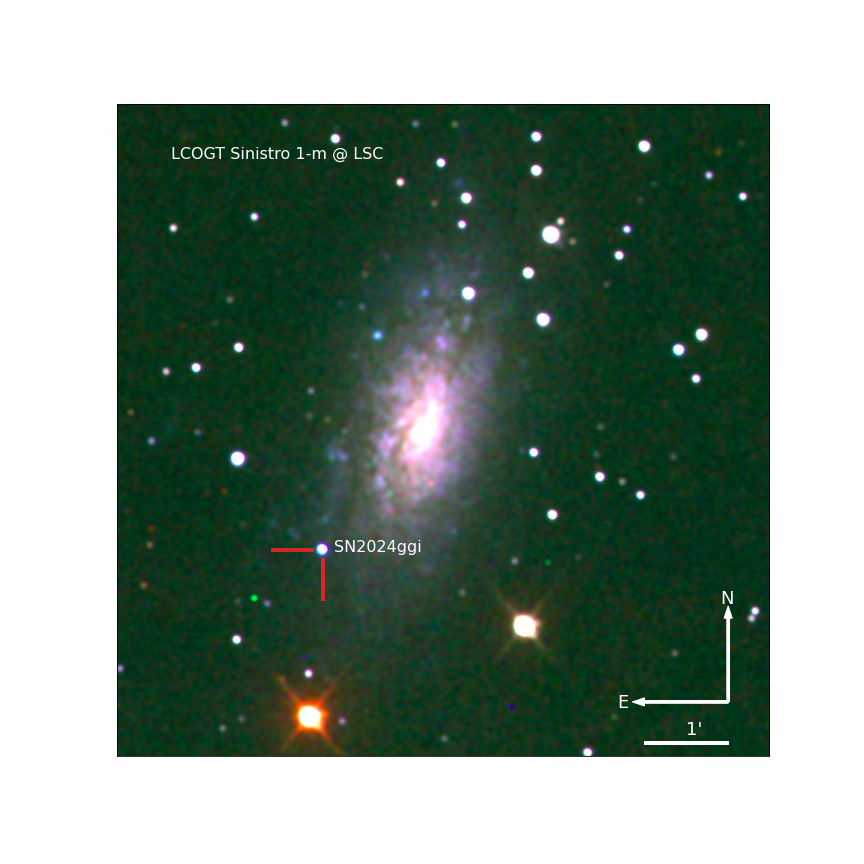}}
\caption{SN 2024ggi on the spiral arms of NGC 3621. The composite $gri$ image was obtained with the LCO $1.0$~m telescope on April 17, 2024. The SN is located at a projected distance of $3.87$~kpc from the galaxy center. \label{fig:ngc3621}} 
\end{figure}

 Figure \ref{fig:ngc3621} shows the position of SN 2024ggi in the spiral arms of NGC 3621, in a composite $gri$ image, obtained with the Las Cumbres Observatory (LCO) $1.0$~m telescope on 2024-04-17.
 The SN is located at at RA = 11:18:22.087, DEC = $-32$:50:15.27.
 Considering the most recent Cepheid distance for NGC 3621 of $7.11$~Mpc \citep{2006ApJS..165..108S}, we estimate the distance of SN 2024ggi to the galaxy center to be of $3.87$~kpc.

\section{Epoch of the first light}\label{app:time}

\begin{figure}[t!]
\centerline{\includegraphics[width=0.45\textwidth]{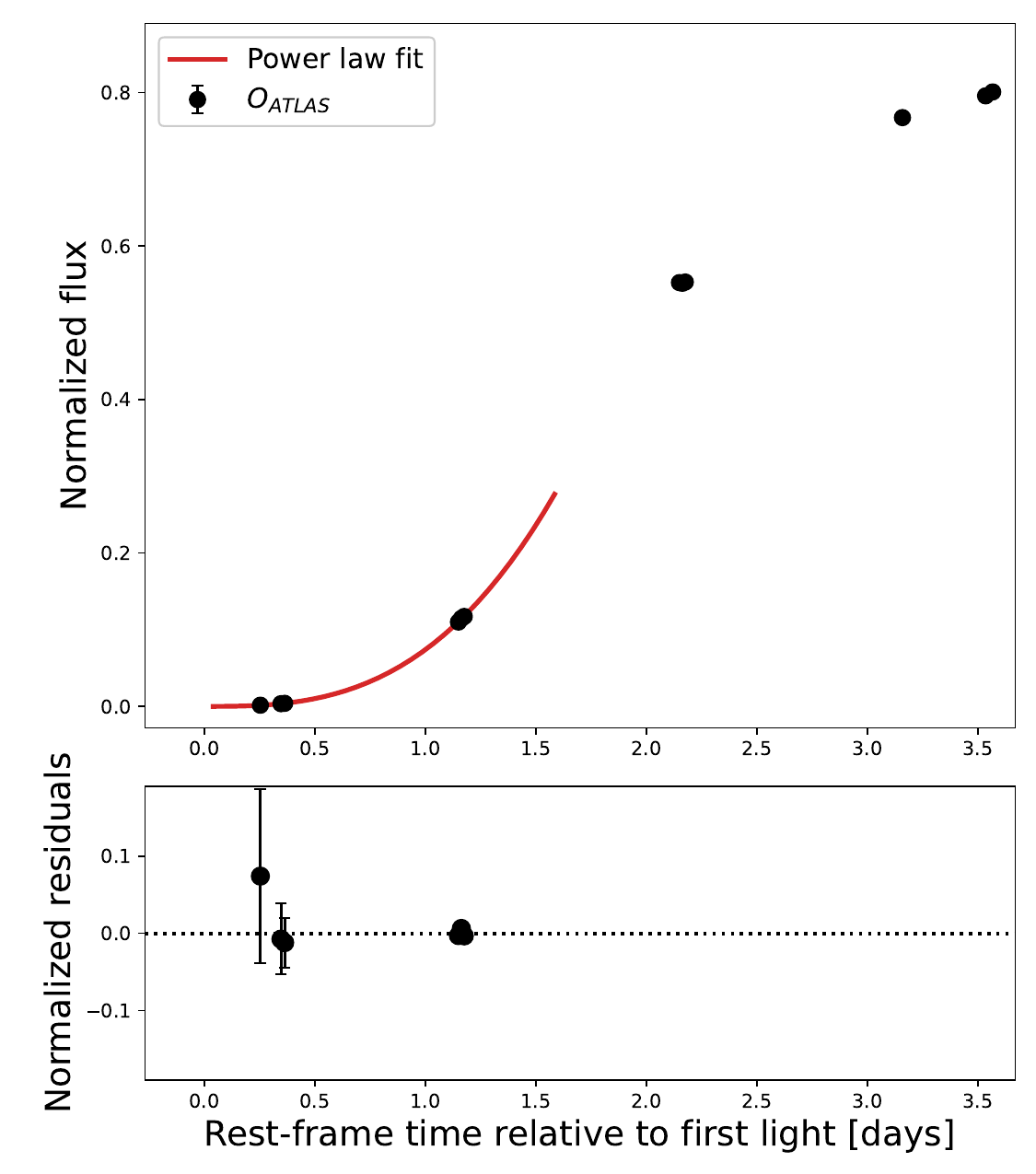}}
\caption{Power law fit to the early ATLAS $o$-band photometry of SN 2024ggi, retrieved from the public ATLAS forced photometry server. The SN rise in the $o$ band is described well by a power law with $\alpha = 2.9 \pm 0.9$. The bottom panel reports the normalized residuals of the fitting. \label{fig:power_law}} 
\end{figure}

The very early discovery of SN 2024ggi by the ATLAS survey and
its subsequent detailed follow-up, allows a 
detailed investigation of its infant phase. We used
the public ATLAS forced photometry obtained from the ATLAS forced photometry
server\footnote{\url{https://fallingstar-data.com/forcedphot/}}
to estimate the epoch of the first SN light. It is assumed that the early evolution of SN 2024ggi  can be described by a power law of the form $F \propto (t-t_{0})^{\alpha}$, where $t_{0}$ is the epoch of the first light and $\alpha$ is the power law exponent \citep[see e.g.,][]{cartier17}. A power law fit to the ATLAS $o_{\mathrm{ATLAS}}$ photometry during the first $24$~h following
the SN discovery (six photometric observations) yields $t_{0} = 60410.89 \pm 0.14$ days (MJD, with $\alpha = 2.9 \pm 0.9$), or $6$~h before the discovery time. The power-law fit describes remarkably well the early rise in the SN brightness, yielding a reduced $\chi^{2}_{\nu} = 0.85$ with an average difference between the model and the observations of \textless $1$\%, as it is shown in Fig. \ref{fig:power_law}. The uncertainty in $t_{0}$ and $\alpha$ parameters was estimated using a jackknife
methodology removing one of the photometric measurements from the fit and
re-fitting, finding an excellent agreement in the parameters each time, in particular at the time of the first light in the $o_{ATLAS}$ filter. 

\section{Extinction and intervening absorbing clouds}\label{app:na}

\begin{figure}[t!]
\centerline{\includegraphics[scale=0.5]{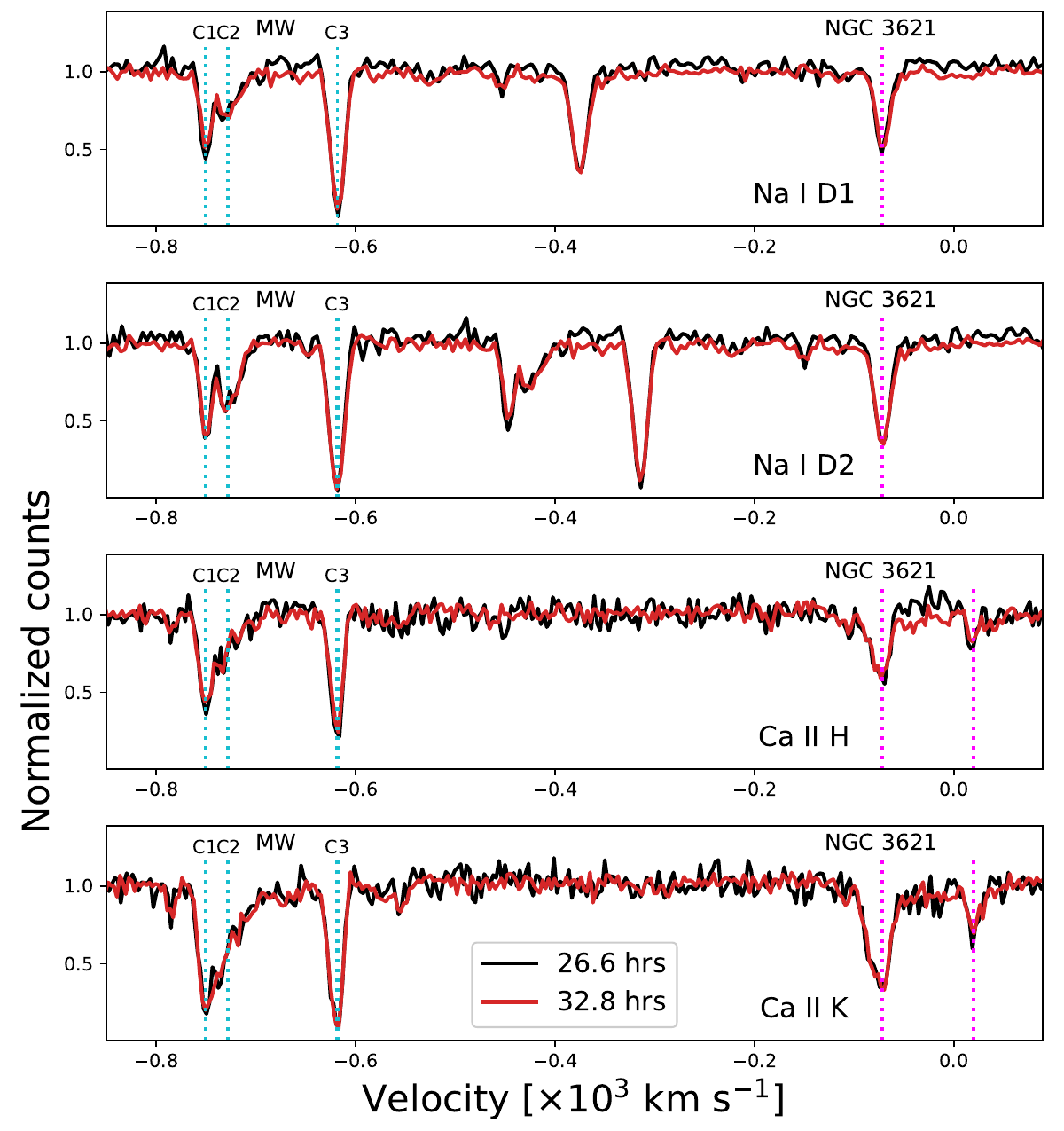}}
\caption{Spectral region around the \ion{Na}{I D1}, \ion{Na}{I D2}, \ion{Ca}{II H}, and \ion{Ca}{II K} absorption features (from top to bottom). The dotted cyan lines mark the position of the Milky Way intervening absorbing clouds, and the dotted pink lines mark the position of the host galaxy features. The velocity scale is relative to the center of NGC 3621. \label{fig:NaI}} 

\end{figure}

The galactic longitude and latitude of SN 2024ggi are $l = 281.24449207$~deg and $b=26.08475353$~deg, respectively. Thus, the line of sight to SN\,2024ggi is not far from the galactic plane. 
The high-resolution spectroscopy of SN\,2024ggi reveals three intervening galactic clouds in the line of sight to the SN, producing absorption lines of \ion{Ca}{II} $H\&K$ and \ion{Na}{I} D1\&D2 lines, at velocities of $-22.0 \pm 0.2$\,km s$^{-1}$, $0.6 \pm 0.1$,km s$^{-1}$ and $+110.4 \pm 0.4$\,km s$^{-1}$ (see Fig. \ref{fig:NaI}). 
In the same way, the spectra reveal \ion{Ca}{II} and \ion{Na}{I} absorption lines produced by intervening clouds in NGC\,3621 at a velocity of $-71.7 \pm 0.4$\,km\,s$^{-1}$ and a weak absorption in the \ion{Ca}{II} H\&K lines at a velocity of $+19.3 \pm 0.9$\,km\,s$^{-1}$ relative to the heliocentric velocity of NGC\,3621 ($v_{\mathrm{helio}} = 730 \pm 2$\,km\,s$^{-1}$).

Using the \citet{poznaski12} relations, we estimate the galactic and the host galaxy reddening. By means of fitting a Gaussian profile to the absorption lines we obtain average \ion{Na}{I} D1+D2 equivalent widths (EWs) of $0.26 \pm 0.01$\,\AA, $0.36 \pm 0.01$\,\AA, and $0.63 \pm 0.01$\,\AA\ for the three intervening galactic clouds and of $0.41 \pm 0.01$\,\AA\ for the \ion{Na}{I} D1+D2 absorption lines in NGC\,3621. Using the \citet{poznaski12} relation for the \ion{Na}{I} D1+D2 EW, these EWs translate to $E(B-V)$ of $0.029 \pm 0.007$\,mag, $ 0.037 \pm 0.009$\,mag, and $0.078 \pm 0.018$ mag for the three clouds in our galaxy, respectively, and of $0.042 \pm 0.008$ mag for NGC\,3621. Very similar results were obtained when the relations of \citet{poznaski12} for the individual EWs of \ion{Na}{I} D2 or \ion{Na}{I} D1 lines were used. Note that \citet{poznaski12} relations were derived using the original extinction maps of \citet{shlegel98}, and the $E(B-V)$ values computed from \citet{poznaski12} relations must be multiplied by $0.86$ to be placed in the more recent recalibration of \citet{schlafly11}. 

The recalibrated galactic extinction maps of \citet{shlegel98} predict a Milky Way extinction of $E(B-V) = 0.0694$ mag \citep{schlafly11} in the line of sight to SN\,2024ggi. This value compares favorably with the combined extinction from clouds 1 ($-22.0 \pm 0.2$\,km s$^{-1}$) and 2 ($0.6 \pm 0.1$,km s$^{-1}$) of $E(B-V) = 0.057 \pm 0.009$ which was scaled by $0.86$ to place this value in the recent recalibration of \citet{schlafly11}, or with the extinction of the third cloud of $E(B-V) = 0.067 \pm 0.015$ \citep[scaled to][]{schlafly11}, but does not seem to account for the total extinction estimated for the three Galactic clouds detected using high-resolution spectroscopy. As noted in the cautionary notes\footnote{\url{https://irsa.ipac.caltech.edu/applications/DUST/docs/background.html##notes}}, the dust extinction map of \citet{shlegel98} is most accurate when a single dust temperature describes the bulk of the dust that is absorbing/scattering background starlight, but closer to the Galactic plane multiple dust clouds can be encountered which emits multiple dust temperature distributions yielding a less accurate estimate on $E(B-V)$, this may explain the discrepancy in the Galactic extinction along the line of sight to SN 2024ggi. In summary we report a total Galactic extinction of $E(B-V)_{\textrm{gal}} = 0.12 \pm 0.02$ mag and a host galaxy extinction of $E(B-V)_{\textrm{host}} = 0.036 \pm 0.007$ mag, producing a total extinction along the line of sight to SN 2024ggi of $E(B-V)_{\textrm{total}} = 0.16 \pm 0.02$ mag.

\end{appendix}
\end{document}